\documentclass[journal=nalefd,manuscript=letter]{achemso}
\usepackage{graphicx}
\usepackage{layout}
\usepackage{dcolumn}
\usepackage{braket}
\usepackage{bm}
\usepackage{color}
\usepackage{verbatim}
\usepackage{txfonts}
\usepackage{textcomp}

\author{A. P. Higginbotham}
\affiliation{Center for Quantum Devices, Niels Bohr Institute, University of Copenhagen, 2100 Copenhagen, Denmark}
\alsoaffiliation{Department of Physics, Harvard University, Cambridge, Massachusetts 02138, USA}

\author{T. W. Larsen}
\affiliation{Center for Quantum Devices, Niels Bohr Institute, University of Copenhagen, 2100 Copenhagen, Denmark}

\author{J. Yao}

\author{H. Yan}

\author{C. M. Lieber}
\affiliation{Department of Chemistry and Chemical Biology, Harvard University, Cambridge, Massachusetts 02138, USA}
\alsoaffiliation{School of Engineering and Applied Sciences, Harvard University, Cambridge, Massachusetts 02138, USA}

\author{C.~ M.~Marcus}

\author{F. Kuemmeth}
\email{kuemmeth@nbi.dk}
\affiliation{Center for Quantum Devices, Niels Bohr Institute, University of Copenhagen, 2100 Copenhagen, Denmark}

\title{Hole Spin Coherence in a Ge/Si Heterostructure Nanowire}

\makeatletter
\let\acs@address@list\relax
\makeatother

\begin{document}

\begin{abstract}
Relaxation and dephasing of hole spins are measured in a gate-defined Ge/Si nanowire double quantum dot using a fast pulsed-gate method and dispersive readout.
An inhomogeneous dephasing time $T_2^* \sim 0.18~\mathrm{\mu s}$ exceeds corresponding measurements in III-V semiconductors by more than an order of magnitude, as expected for predominately nuclear-spin-free materials.
Dephasing is observed to be exponential in time, indicating the presence of a broadband noise source, rather than Gaussian, previously seen in systems with nuclear-spin-dominated dephasing.
\end{abstract}

\maketitle


\noindent
\textbf{Competing financial interests:} The authors declare no competing financial interests.
\pagebreak

Realizing qubits that simultaneously provide long coherence times and fast control is a key challenge for quantum information processing.
Spins in III-V semiconductor quantum dots can be electrically manipulated, but lose coherence due to interactions with nuclear spins \cite{Petta:2005kna,Koppens:2006kz,Hanson:2007eg}.
While dynamical decoupling and feedback have greatly improved coherence in III-V qubits \cite{Bluhm:2010fj,Medford:2012fy,Bluhm:2010et}, the simple approach of eliminating nuclear spins using group IV materials remains favorable.
Carbon nanotubes have been investigated for this application \cite{Churchill:2009fy,Churchill:2009fs,Pei:2012jn,Laird:2013ez}, but are difficult to work with due to uncontrolled, chirality-dependent electronic properties. So far, coherence has not been improved over III-V spin qubits.

Si devices have shown improved coherence for gate-defined electron quantum dots \cite{Goswami:2006bb,Shaji:2008hq,Maune:2012iu,Shi:2013uf}, and for electron and nuclear spins of phosphorous donors \cite{Pla:2012jj,Pla:2013io,Dehollain:2014wc,Muhonen:2014tb}.
The Ge/Si core/shell heterostructured nanowire is an example of a predominantly zero-nuclear-spin system that is particularly tunable and scalable \cite{Lu:2005jx,Xiang:2006gi,Yan:2011cm,Yao:2013eo,Yao:uh}.
Holes in Ge/Si nanowires exhibit large spin-orbit coupling \cite{Hao:2010di,Kloeffel:2011bg,Higginbotham:2014ui}, a useful resource for fast, all-electrical control of single spins \cite{Nowack:2007du,NadjPerge:2010kw,Nowack:2011cu,vandenBerg:2013dk,Laird:2013ez}.
Moving to holes should also improve coherence because the contact hyperfine interaction, though strong for electrons associated with $s$-orbitals, is absent for holes associated with $p$-orbitals \cite{Fischer:2008hh}.
Indeed, a suppression of electron-nuclear coupling in hole conductors was recently demonstrated in InSb \cite{Pribiag:2013if}.

Here, we measure spin coherence times of gate-confined hole spins in a Ge/Si nanowire double quantum dot using high bandwidth electrical control and read out of the spin state.
We find inhomogeneous dephasing times $T_2^*$ up to $0.18~\mathrm{\mu s}$, twenty times longer than in III-V semiconductors. 
This timescale is consistent with dephasing due to sparse $^{73}$Ge nuclear spins.
The observed exponential coherence decay suggests a dephasing source with high-frequency spectral content, and we discuss a few candidate mechanisms.
These results pave the way towards improved spin-orbit qubits and strong spin-cavity coupling in circuit quantum electrodynamics \cite{Kloeffel:2013iv}.

Ge/Si core/shell nanowires host a tunable hole gas in the Ge core [Fig.~1(a)] with typical mobility $\mu \sim 1000~\mathrm{cm^2/(V \cdot s)}$.
In the presence of realistic external electric fields, the 1D hole gas is expected to occupy a single Rashba-split subband with $\sim1~\mathrm{meV}$ spin-orbit splitting, based on theory \cite{Kloeffel:2011bg} and previous experiments \cite{Hao:2010di,Higginbotham:2014ui}.
Fabrication of double dots with discrete hole states, and measurements of spin relaxation have been reported \cite{Hu:2007hu,Hu:2011ic}.

The device, diagrammed in Fig.~1(b), is fabricated on a lightly doped Si substrate.
The substrate, insulating at $T<10~\mathrm{K}$,  is covered with $\mathrm{HfO_2}$ using atomic layer deposition.
Nanowires are deposited from methanol solution and contacted by evaporating Al following a buffered hydrofluoric acid dip.
A second layer of $\mathrm{HfO_2}$ covers the wire, and Cr/Au electrostatic gates are placed on top.
These gates tune the hole density along the length of the wire. 
All data are obtained at temperature $T < 100\mathrm{~mK}$ in a dilution refrigerator with external magnetic field $B = 0$, unless otherwise noted.

\begin{figure}[b]
	\includegraphics{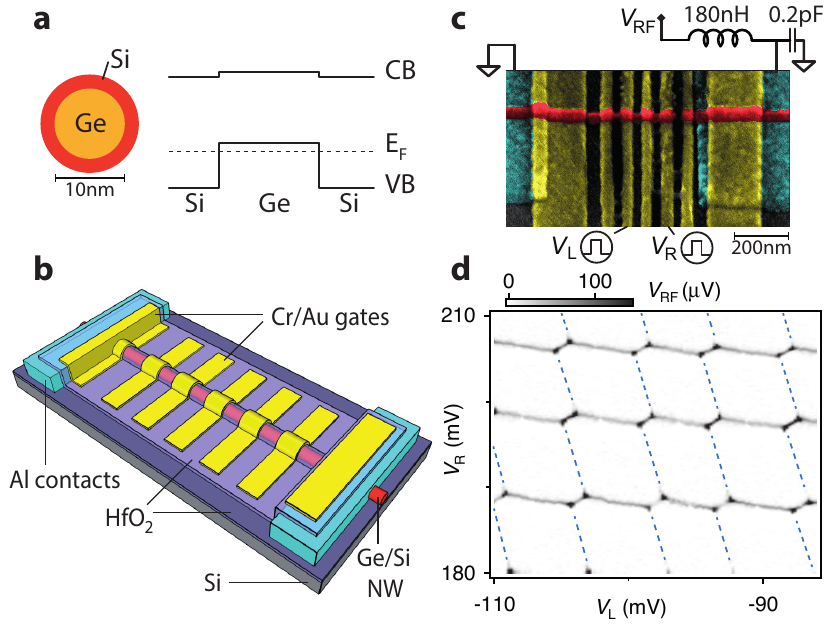}
	\caption{
	Ge/Si double quantum dot device.
	\label{fig:1} \textbf{a}, Cross section and energy diagram of conduction band (CB) and valence band (VB). The quantum well supporting the hole gas forms in the VB of Ge.
	\textbf{b}, Device schematic.
	\textbf{c},  False color scanning electron micrograph.
	High-bandwidth plunger gates $V_\mathrm{L}$ and $V_\mathrm{R}$ are labeled. 
	$V_\mathrm{RF}$ is reflected from the LC circuit attached on the right lead.
	\textbf{d},~ Demodulated $V_\mathrm{RF}$ versus $V_\mathrm{L}$ and $V_\mathrm{R}$ at $B=1$ T.
	Negatively sloped gray lines correspond to single-hole transfers between the right dot and right lead.
	Positive slopes are due to hole transfers directly between dots.
	Guides to the eye (dashed lines) indicate hole transfers between the left dot and lead, too faint to be visible in the data because the resonator is on the right side.
	}
\end{figure}

Gate voltages are tuned to form a double quantum dot in the nanowire with control over charge occupancy and tunnel rates.
High-bandwidth ($400~\mathrm{MHz}$) plunger gates $V_\mathrm{L}$ and $V_\mathrm{R}$, labeled in Fig.~1(c), control hole occupation in the left and right dots.
The readout circuit is formed by wire bonding a 180~nH inductor directly to the source electrode of the device.
Combined with a total parasitic capacitance of $0.2~\mathrm{pF}$, this forms an LC resonance at $830~\mathrm{MHz}$ with bandwidth $15~\mathrm{MHz}$.
Tunneling of holes between dots or between the right dot and lead results in a capacitive load on the readout circuit, shifting its resonant frequency. \cite{Petersson:2010tq,Jung:2012dz}.
The circuit response is monitored by applying near-resonant excitation to the readout circuit and recording changes in the reflected voltage, $V_\mathrm{RF}$, after amplification at $T=4~\mathrm{K}$ and demodulation using a  $90^{\circ}$ power splitter and two mixers  at room temperature 
\bibnote{ S. Weinreb LNA SN68. Minicircuits ZP-2MH mixers. Tektronix AWG5014 waveform generator used on $V_\mathrm{L}$ and $V_\mathrm{R}$. Coilcraft 0603CS chip inductor.}.

The charge stability diagram of the double dot is measured by monitoring $V_\mathrm{RF}$ at fixed frequency while slowly sweeping $V_\mathrm{L}$ and $V_\mathrm{R}$ [Fig.~1(d)].
Lines are observed whenever single holes are transferred to or from the right dot.
Transitions between the left dot and left lead are below the noise floor (not visible) because the LC circuit is attached to the right lead.
Enhanced signal is observed at the triple points, where tunneling is energetically allowed across the entire device.
The observed ``honeycomb" pattern is consistent with that of a capacitively-coupled double quantum dot \cite{vanderWiel:2002gr}.
The charging energies for the left and right dots are estimated $1.7~\mathrm{meV}$ and $2.7~\mathrm{meV}$ from Fig.~1(d), using a plunger lever arm of $0.3$~eV/V, determined from finite bias measurements on similar devices \cite{Hu:2007hu}.
The few-hole regime was accessible only in the right dot, identified by an increase in charging energy.
Based on the location of the few-hole regime in the right dot, we estimate the left and right hole occupations to be 70 and 10 at the studied tuning.
We found that operating in the many-hole regime improved device stability, facilitating gate tuning and readout.
We do not know if this affects the quality of the qubit, as recently found for electron spins in GaAs \cite{Higginbotham:2014cg}.

The spin state of the double dot is read out by mapping it onto a charge state using the Pauli blockade pulse sequence diagrammed in Fig.~2.
At the points E1 and E2 (``empty'') the double dot is in the (m+2,~n+1) charge state, assuming that m (n) paired holes occupy lower orbitals in the left (right) dot. 
Pulsing to P (``prepare'') in (m+1,~n+1) discards one hole from the left dot, leaving the spin state of the double dot in a random mixture of singlet and triplet states.
Moving to M (``measure'') adjusts the energy detuning between the dots, making interdot tunneling favorable.
When M is located at zero detuning, $\varepsilon = 0$, tunneling is allowed for singlet but Pauli-blocked for triplet states.
When M is at the singlet-triplet splitting, $\varepsilon = \Delta_\mathrm{ST}$, triplet states can tunnel.
The location of the interdot charge transition therefore reads out the spin state of the double dot.
We expect this picture to be valid for multi-hole dots with an effective spin-$\frac{1}{2}$ ground state \cite{Hu:2011ic,Vorojtsov:2004ho,Johnson:2005cv,Higginbotham:2014cg}.
We use singlet-triplet terminology for clarity, but note that strong spin-orbit coupling changes the spin makeup of the blockaded states without destroying Pauli blockade \cite{Danon:2009id}.

The fast pulse sequence $\mathrm{E1}{\rightarrow}\mathrm{E2}{\rightarrow}\mathrm{P}{\rightarrow}\mathrm{M}{\rightarrow}\mathrm{E1}$ is repeated continuously while rastering the position of $\mathrm{M}=(V_\mathrm{L},V_\mathrm{R})$ near the (m+1, n+1)-(m+2, n) charge transition (Fig.~2).
The RF carrier is applied only at the measurement point, M.
As shown in Fig.~1(d), features with negative slope are observed corresponding to transitions across the right barrier. 
We interpret the weak interdot transition at zero detuning accompanied by a relatively strong interdot feature at large detuning as Pauli blockade of the ground-state interdot transition ($\varepsilon = 0$), and lifting of blockade at the singlet-triplet splitting ($\varepsilon = \Delta_\mathrm{ST}$).
The strength of the $\varepsilon = 0$ interdot transition thus measures the probability of loading a singlet at point P, while the strength at $\varepsilon = \Delta_\mathrm{ST}$ measures the probability of loading a triplet.
As a control, the Pauli blockade pulse sequence was run in the opposite direction, and no blockade was observed (see Supporting Information).

\begin{figure}
	\includegraphics[width=8.46cm]{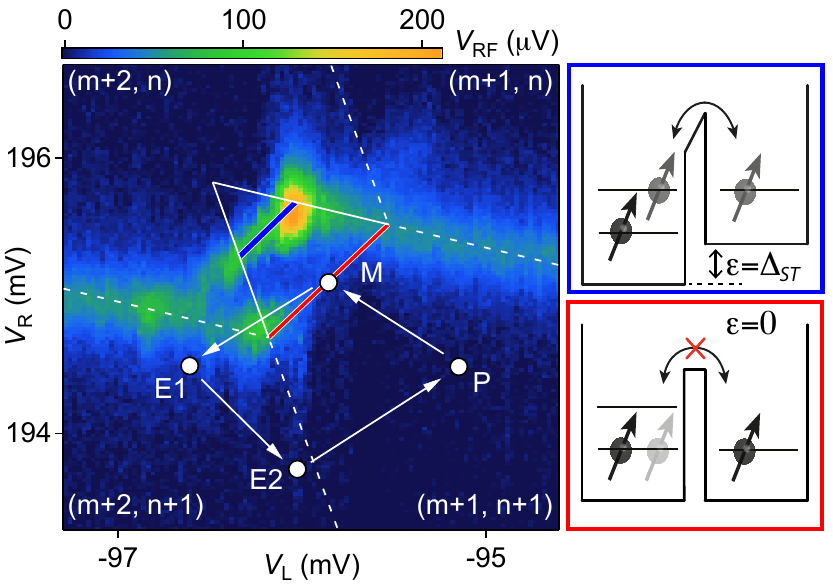}
	\caption{\label{fig:2}
	Spin readout using Pauli blockade.
	$V_\mathrm{RF}$ at the measurement point $\mathrm{M}=(V_\mathrm{L},V_\mathrm{R})$ of the pulse sequence indicated by white arrows.
	Dashed lines estimate changes in double dot hole occupancy (m,~n), where m (n) denotes the occupancy of the left (right) dot.
	Large solid triangle outlines the region over which direct interdot charge transitions occur.
	The interdot transition at $\varepsilon=0$ (marked by a red line) is weak due to Pauli blockade of triplet states, illustrated in the red diagram.
	The interdot transition at $\varepsilon=\Delta_\mathrm{ST}$ (marked by a blue line) is strong due to tunneling of triplet states, illustrated in the blue diagram.
	}
\end{figure}

\begin{figure}
	\includegraphics{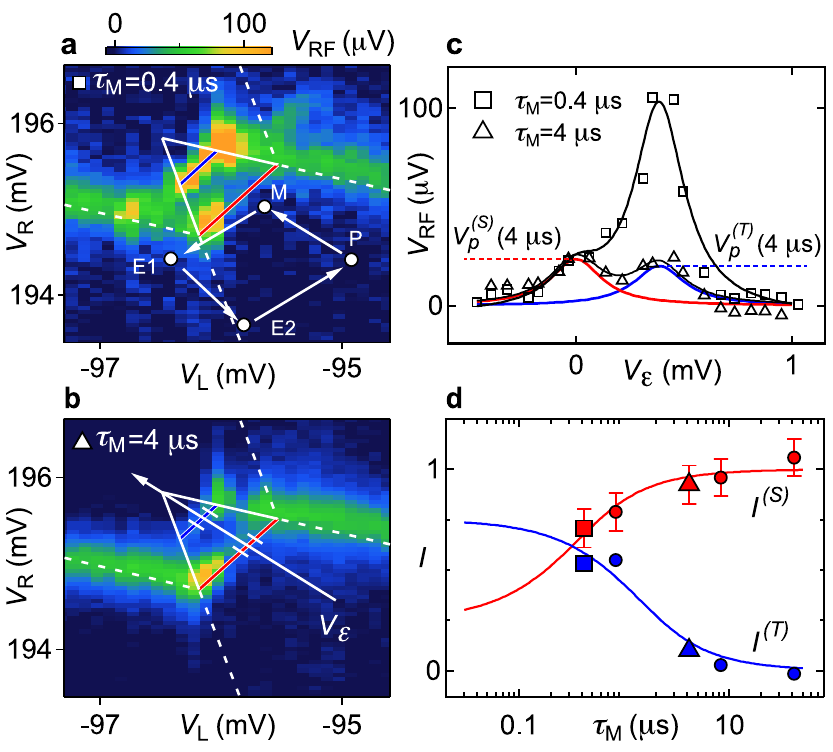}
	\caption{\label{fig:3}
	Spin relaxation.
	\textbf{a}, $V_\mathrm{RF}$ at the measurement point $\mathrm{M=(V_\mathrm{L},V_\mathrm{R})}$ of $T_1$ pulse sequence (arrows). The dwell time at M is $\tau_\mathrm{M}=0.4~\mathrm{\mu s}$.
	\textbf{b}, Same as (\textbf{a}), but with $\tau_\mathrm{M}=4~\mathrm{\mu s}$.
	\textbf{c}, Cuts along the $V_\varepsilon$ region indicated in (\textbf{b}) for $\tau_\mathrm{M}=0.4~\mathrm{\mu s}$ ($\Box$) and $\tau_\mathrm{M}=4~\mathrm{\mu s}$ ($\triangle$).
	Each cut is fit with the sum of two Lorentzians, the left of height $V_p^{(S)}$ and right of height  $V_p^{(T)}$.
	The center of the left Lorentzian defines zero detuning, $V_\varepsilon=0$.
	\textbf{d}, Readout visibility $I^{(S,T)} = V_p^{(S,T)}/V_0^{(S,T)}$ as a function of $\tau_\mathrm{M}$.
	Fits are to Eqs. (1,2) and have characteristic decay times $T_1^{(S)} = 200~\mathrm{ns}$ and $T_1^{(T)} = 800~\mathrm{ns}$ for singlet and triplet states.
	Normalization factors are $V_0^{(S)}=25~\mathrm{\mu eV}$ and $V_0^{(T)}=200~\mathrm{\mu eV}$.
	}
\end{figure}

Spin relaxation is measured by varying the dwell time $\tau_\mathrm{M}$ at the measurement point for the counterclockwise Pauli-blockade sequence. As $\tau_\mathrm{M}$ increases the triplet transition weakens and the singlet transition strengthens [Fig.~3(a,b)] due to triplet-to-singlet spin relaxation.
Note that these relaxation processes have different charge characters at different measurement points.
For example, at $\varepsilon = 0$ the initial charge state is (m+1, n+1), whereas at $\varepsilon = \Delta_\mathrm{ST}$ the initial charge state is hybridized with (m+2, n).

The $T_1$ spin relaxation time is measured by analyzing a cut along the $V_\varepsilon$ axis [shown in Fig.~3(b)] and varying $\tau_\mathrm{M}$.
For each $\tau_\mathrm{M}$, the cut is fit to the sum of two Lorentzians with equal widths and constant spacing.
The heights are $V_p^{(T)}$ for the triplet peak and $V_p^{(S)}$ for the singlet peak.
Two example cuts are shown in Fig.~3(c), along with fits to exponential forms,
\begin{eqnarray}
	V_p^{(S)}(\tau_\mathrm{M}) &= \frac{1}{4} V_0^{(S)}\left[4 - 3 p\left(\tau_\mathrm{M},T_1^{(S)}\right)\right], \\
	V_p^{(T)}(\tau_\mathrm{M}) &= \frac{3}{4}V_0^{(T)} p\left(\tau_\mathrm{M},T_1^{(T)}\right),
\end{eqnarray}
where $p(\tau_\mathrm{M},T_1) = (1 / \tau_\mathrm{M} ) \int_0^{\tau_\mathrm{M}} e^{-t/T_1}\mathrm{d}t$ is the exponential decay averaged over the measurement time.
Figure~3(d) plots the readout visibility, $I^{(S,T)}=~V_p^{(S,T)}/V_0^{(S,T)}$.
The extracted relaxation time is $T_1^{(T)} = 800~\mathrm{ns}$ at the triplet position [blue line in Figs.~3(a,b)], and $T_1^{(S)} = 200~\mathrm{ns}$ at the singlet position [red line in Figs.~3(a,b)].
We note that these spin relaxation times are three orders of magnitude shorter than those previously measured in a similar device in a more isolated gate configuration and away from interdot transitions \cite{Hu:2011ic}.
Detuning dependence of spin relaxation has been observed previously and attributed to detuning-dependent coupling to the leads as well as hyperfine effects (presumably the former dominate here) \cite{Johnson:2005ks,Petersson:2013cv}.
Relaxation due to the spin-orbit interaction is expected to take microseconds or longer \cite{Maier:2013gg}.
The difference between $V_0^{(S)}$ and $V_0^{(T)}$ can possibly be attributed to differences in singlet-singlet and triplet-triplet tunnel couplings or enhanced coupling near the edges of the pulse triangle.
The separation between Lorentzian peaks by $0.38~\mathrm{mV}$ can be interpreted as $\Delta_\mathrm{ST} = 160~\mathrm{\mu eV}$, using a plunger lever arms of $0.3$~eV/V.

\begin{figure}
	\includegraphics{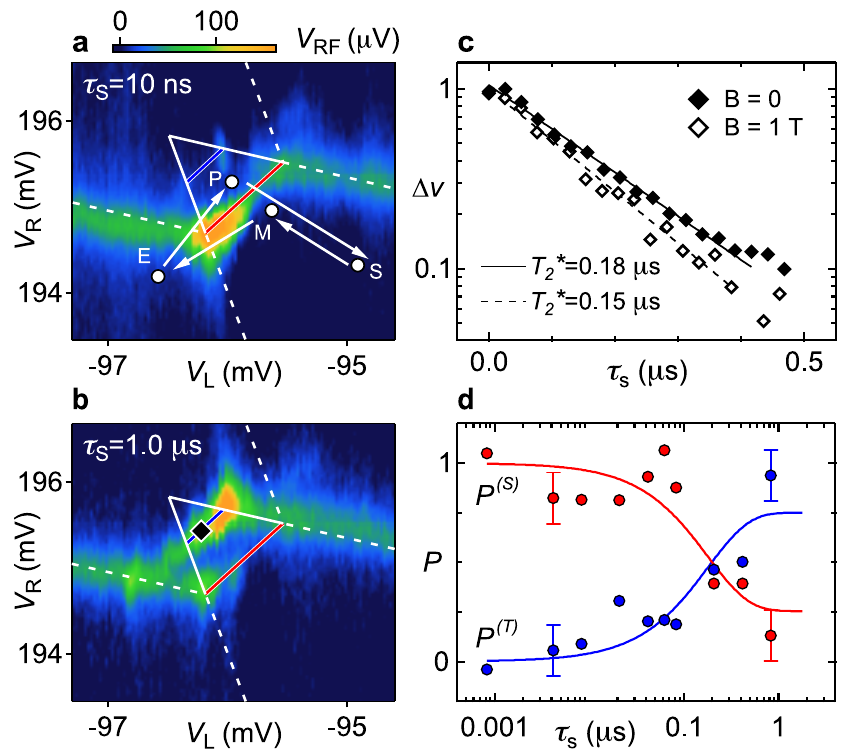}
	\caption{\label{fig:4}
	Spin dephasing.
	\textbf{a}, $V_\mathrm{RF}$ at the measurement point $\mathrm{M}=(V_\mathrm{L}, V_\mathrm{R})$ of $T_2^*$ pulse sequence (arrows). 
	The dwell time at S is $\tau_\mathrm{S}=10~\mathrm{ns}$.
	\textbf{b}, Same as (\textbf{a}), but with $\tau_\mathrm{s}=1~\mathrm{\mu s}$.
	\textbf{c}, Normalized differential voltage at the triplet line $\Delta v \equiv [ V( \tau_\mathrm{S} ) - V_\infty ] / [ V( 0 ) - V_\infty ]$ as a function of $\tau_\mathrm{S}$.
	The $B = 0~T$ data are measured at $(V_\mathrm{L}, V_\mathrm{R})$ indicated in (\textbf{b}), yielding a $T_2^*$ dephasing time of $0.18~\mathrm{\mu s}$.
	The $B = 1~$T data are obtained at a different dot occupancy and tuning using the same method, yielding $T_2^*=0.15~\mathrm{\mu s}$.
	The normalization factor is $V_\mathrm{RF}( 0 ){-}V_\infty=35~\mathrm{\mu V}$.
	Solid and dashed lines are fits to exponentials.
	\textbf{d}, Probability $P^{(S,T)} = V_p^{(S,T)}/V_0^{(S,T)}$ obtained from data as in (\textbf{a},\textbf{b}), analyzed as in Fig.~3(c).
	Fits are to Eqs.~(3,4) with $T_2^*=0.18~\mathrm{\mu s}$ fixed from (\textbf{c}).
	Normalization factors are $V_0^{(S)}=60~\mathrm{\mu eV}$ and $V_0^{(T)}=130~\mathrm{\mu eV}$.
	}
\end{figure}

To investigate spin dephasing, an alternate pulse sequence is used that first initializes the system into a singlet state in (m+2,~n) at point P, then separates to point S (``separate'') in (m+1,~n+1) for a time $\tau_\mathrm{S}$ [Fig.~4(a)].
The spin state of the double dot is measured at M by pulsing back towards (m+2,~n).
For short $\tau_\mathrm{S}$ [Fig.~4(a)] a strong singlet return feature is observed, consistent with negligible spin dephasing.
For long $\tau_\mathrm{S}$ [Fig.~4(b)], a strong triplet return feature is observed, consistent with complete spin dephasing.

The $T_2^*$ dephasing time is found by measuring $V_\mathrm{RF}(\tau_\mathrm{S})$ at the triplet transition, and plotting the normalized differential voltage  $\Delta v \equiv [ V_\mathrm{RF}( \tau_\mathrm{S} ){-}V_\infty ] / [ V_\mathrm{RF}( 0 ){-}V_\infty ]$ as a function of separation time [Fig.~4(c)].
Here, $V_\infty \equiv V_\mathrm{RF}( 500~\mathrm{ns} )$ is the demodulated voltage for a pulse sequence with long dephasing time. 
The quantity $[ V_\mathrm{RF}( \tau_\mathrm{S} ){-}V_\infty ]$ is directly measured by alternating between the $T_2^*$ sequence and a reference sequence with long dephasing time, and feeding the demodulated voltage into a lock-in amplifier.
Fitting the $B = 0$ data to $\rm{exp}[-(\tau_\mathrm{S}/T_2^*)^\alpha]$ yields $\alpha = 1.1 \pm 0.1$. 
Figure~4(c) shows exponential fits ($\alpha = 1$) for both data sets.
The $B = 0$ data decays exponentially on a timescale $T_2^*=0.18~\mathrm{\mu s}$.
Data acquired at $B = 1~\mathrm{T}$ at a different double-dot occupation give a similar timescale and functional form.

Although this timescale is approaching the limit expected for dephasing due to random Zeeman gradients from sparse $^{73}$Ge nuclear spins (see Supporting Information), the observed exponential loss of coherence is by and large unexpected for nuclei.
A low-frequency-dominated nuclear bath is expected to yield a Gaussian fall-off of coherence with time \cite{Coish:839687}, in contrast to the observed exponential dependence, which instead indicates a rapidly varying bath \cite{Cywinski:2008iy}.
Nuclei can produce high-bandwidth noise in the presence of spatially varying effective magnetic fields, for example due to inhomogeneous strain-induced quadrupolar interactions \cite{Beaudoin:2013bm}. 
The similarity of data at $B = 0$ and $B =1$~T in Fig.~4(c), however, would indicate an unusually large energy-scale for nuclear effects.
Electrical noise, most likely from the sample itself, combined with spin-orbit coupling is a plausible alternative.
For electrons, the ubiquitous $1/f$ electrical noise alone does not result in pure dephasing \cite{Huang:2013wz}, but can add high-frequency noise to the low-frequency contribution from the nuclear bath.
It is conceivable that the behavior is different for holes, but this has not been studied to our knowledge.
The relative importance of nuclei versus electrical noise could be quantified in future experiments by studying spin coherence in isotopically pure Ge/Si nanowires.

Cuts along the $V_\varepsilon$ axis in Fig.~4(b) as a function of $\tau_\mathrm{S}$ provide a second method for obtaining $T_2^*$, following analysis along the lines of Fig.~3(c).
The resulting probability $P^{(S,T)}  = V_p^{(S,T)}/V_0^{(S,T)}$ versus $\tau_\mathrm{S}$ is shown in Fig.~4(d), along with  exponential curves
\begin{eqnarray}
	V_p^{(S)}(\tau_\mathrm{S}) &= P_\infty V_0^{(S)}\left[1 - (1-1/P_\infty)e^{-\tau_\mathrm{S}/T_2^*}\right], \\
	V_p^{(T)}(\tau_\mathrm{S}) &= (1-P_\infty ) V_0^{(T)}\left[1 - e^{-\tau_\mathrm{S}/T_2^*}\right],
\end{eqnarray}
using $T_2^*=0.18 \mathrm{~\mu s}$, with $P_\infty$ and $V_0^{(S,T)}$ as fit parameters.
Depending on the nature of the dephasing, the singlet probability settling value, $P_\infty$, is expected to range from $1/3$ for quasi-static Zeeman gradients to $1/4$ for rapidly varying baths \cite{Schulten:1978jm,Merkulov:2002dz,Zhang:2006bd}.
We find $P_\infty = 0.25 \pm 0.08$.
Equations (3,4) do not take into account spin relaxation at the measurement point, meaning that the fitted $P_\infty$ systematically overestimates the true settling value \bibnote{We do not correct for $T_1$ effects in Eqs. (3,4), as $V_\mathrm{max}^{(S,T)}$ differed significantly from those observed in Fig.~3.}.
Therefore, we conclude that the data weakly support $P_\infty=1/4$ rather than $P_\infty=1/3$, consistent with our inference of a rapidly varying bath.

Unexplained high-frequency noise has recently been observed in other strong spin-orbit systems, such as InAs nanowires \cite{NadjPerge:2010kw}, InSb nanowires \cite{vandenBerg:2013dk}, and carbon nanotubes \cite{Laird:2013ez}.
In these systems slowly varying nuclear effects were removed using dynamical decoupling, revealing the presence of unexplained high-frequency noise. 
In our system the effect of nuclei is reduced by the choice of material, and an unexplained high-frequency noise source appears directly in the $T_2^*$.
These similarities suggest the existence of a shared dephasing mechanism  that involves spin-orbit coupling.

Future qubits based on Ge/Si wires could be coupled capacitively \cite{Trifunovic:2012iq,Shulman:2012fka} or through a cavity using circuit quantum electrodynamics \cite{Kloeffel:2013iv,Petersson:2013cv}.
In the latter case, the long dephasing times measured here suggest that the strong coupling regime may be accessible.

\begin{acknowledgement}
We acknowledge helpful discussions with F\'{e}lix Beaudoin, William Coish, Jeroen Danon, Xuedong Hu, Christoph Kloeffel, Daniel Loss, Franziska Maier and Mark Rudner, and experimental assistance from Patrick Herring.
Funding from the Department of Energy, Office of Science \& SCGF, the EC FP7-ICT project SiSPIN \textnumero\ 323841, and the Danish National Research Foundation is acknowledged.
\end{acknowledgement}

\begin{suppinfo}
Supporting information is provided on acquisition method for Figure 1d, image analysis methods, clockwise $T_1$ pulse sequence (control experiment), and theoretical estimate of $T_2^*$ timescale for Ge/Si nanowire.
\end{suppinfo}

\providecommand*\mcitethebibliography{\thebibliography}
\csname @ifundefined\endcsname{endmcitethebibliography}
  {\let\endmcitethebibliography\endthebibliography}{}


\begin{mcitethebibliography}{57}
\providecommand*\natexlab[1]{#1}
\providecommand*\mciteSetBstSublistMode[1]{}
\providecommand*\mciteSetBstMaxWidthForm[2]{}
\providecommand*\mciteBstWouldAddEndPuncttrue
  {\def\EndOfBibitem{\unskip.}}
\providecommand*\mciteBstWouldAddEndPunctfalse
  {\let\EndOfBibitem\relax}
\providecommand*\mciteSetBstMidEndSepPunct[3]{}
\providecommand*\mciteSetBstSublistLabelBeginEnd[3]{}
\providecommand*\EndOfBibitem{}
\mciteSetBstSublistMode{f}
\mciteSetBstMaxWidthForm{subitem}{(\alph{mcitesubitemcount})}
\mciteSetBstSublistLabelBeginEnd
  {\mcitemaxwidthsubitemform\space}
  {\relax}
  {\relax}

\bibitem[Petta et~al.(2005)Petta, Johnson, Taylor, Laird, Yacoby, Lukin,
  Marcus, Hanson, and Gossard]{Petta:2005kna}
Petta,~J.~R.; Johnson,~A.~C.; Taylor,~J.~M.; Laird,~E.~A.; Yacoby,~A.;
  Lukin,~M.~D.; Marcus,~C.~M.; Hanson,~M.~P.; Gossard,~A.~C. \emph{Science}
  \textbf{2005}, \emph{309}, 2180--2184\relax
\mciteBstWouldAddEndPuncttrue
\mciteSetBstMidEndSepPunct{\mcitedefaultmidpunct}
{\mcitedefaultendpunct}{\mcitedefaultseppunct}\relax
\EndOfBibitem
\bibitem[Koppens et~al.(2006)Koppens, Buizert, Tielrooij, Vink, Nowack,
  Meunier, Kouwenhoven, and Vandersypen]{Koppens:2006kz}
Koppens,~F.; Buizert,~C.; Tielrooij,~K.-J.; Vink,~I.~T.; Nowack,~K.~C.;
  Meunier,~T.; Kouwenhoven,~L.~P.; Vandersypen,~L. \emph{Nature} \textbf{2006},
  \emph{442}, 766--771\relax
\mciteBstWouldAddEndPuncttrue
\mciteSetBstMidEndSepPunct{\mcitedefaultmidpunct}
{\mcitedefaultendpunct}{\mcitedefaultseppunct}\relax
\EndOfBibitem
\bibitem[Hanson et~al.(2007)Hanson, Kouwenhoven, Petta, Tarucha, and
  Vandersypen]{Hanson:2007eg}
Hanson,~R.; Kouwenhoven,~L.~P.; Petta,~J.~R.; Tarucha,~S.; Vandersypen,~L.
  M.~K. \emph{Rev. Mod. Phys.} \textbf{2007}, \emph{79}, 1217\relax
\mciteBstWouldAddEndPuncttrue
\mciteSetBstMidEndSepPunct{\mcitedefaultmidpunct}
{\mcitedefaultendpunct}{\mcitedefaultseppunct}\relax
\EndOfBibitem
\bibitem[Bluhm et~al.(2011)Bluhm, Foletti, Neder, Rudner, Mahalu, Umansky, and
  Yacoby]{Bluhm:2010fj}
Bluhm,~H.; Foletti,~S.; Neder,~I.; Rudner,~M.; Mahalu,~D.; Umansky,~V.;
  Yacoby,~A. \emph{Nat. Phys.} \textbf{2011}, \emph{7}, 109--113\relax
\mciteBstWouldAddEndPuncttrue
\mciteSetBstMidEndSepPunct{\mcitedefaultmidpunct}
{\mcitedefaultendpunct}{\mcitedefaultseppunct}\relax
\EndOfBibitem
\bibitem[Medford et~al.(2012)Medford, Cywi{\'n}ski, Barthel, Marcus, Hanson,
  and Gossard]{Medford:2012fy}
Medford,~J.; Cywi{\'n}ski,~{\L}.; Barthel,~C.; Marcus,~C.~M.; Hanson,~M.~P.;
  Gossard,~A.~C. \emph{Phys. Rev. Lett.} \textbf{2012}, \emph{108},
  086802\relax
\mciteBstWouldAddEndPuncttrue
\mciteSetBstMidEndSepPunct{\mcitedefaultmidpunct}
{\mcitedefaultendpunct}{\mcitedefaultseppunct}\relax
\EndOfBibitem
\bibitem[Bluhm et~al.(2010)Bluhm, Foletti, Mahalu, Umansky, and
  Yacoby]{Bluhm:2010et}
Bluhm,~H.; Foletti,~S.; Mahalu,~D.; Umansky,~V.; Yacoby,~A. \emph{Phys. Rev.
  Lett.} \textbf{2010}, \emph{105}, 216803\relax
\mciteBstWouldAddEndPuncttrue
\mciteSetBstMidEndSepPunct{\mcitedefaultmidpunct}
{\mcitedefaultendpunct}{\mcitedefaultseppunct}\relax
\EndOfBibitem
\bibitem[Churchill et~al.(2009)Churchill, Kuemmeth, Harlow, Bestwick, Rashba,
  Flensberg, Stwertka, Taychatanapat, Watson, and Marcus]{Churchill:2009fy}
Churchill,~H. O.~H.; Kuemmeth,~F.; Harlow,~J.~W.; Bestwick,~A.~J.;
  Rashba,~E.~I.; Flensberg,~K.; Stwertka,~C.~H.; Taychatanapat,~T.;
  Watson,~S.~K.; Marcus,~C.~M. \emph{Phys. Rev. Lett.} \textbf{2009},
  \emph{102}, 166802\relax
\mciteBstWouldAddEndPuncttrue
\mciteSetBstMidEndSepPunct{\mcitedefaultmidpunct}
{\mcitedefaultendpunct}{\mcitedefaultseppunct}\relax
\EndOfBibitem
\bibitem[Churchill et~al.(2009)Churchill, Bestwick, Harlow, Kuemmeth, Marcos,
  Stwertka, Watson, and Marcus]{Churchill:2009fs}
Churchill,~H. O.~H.; Bestwick,~A.~J.; Harlow,~J.~W.; Kuemmeth,~F.; Marcos,~D.;
  Stwertka,~C.~H.; Watson,~S.~K.; Marcus,~C.~M. \emph{Nat. Phys.}
  \textbf{2009}, \emph{5}, 321--326\relax
\mciteBstWouldAddEndPuncttrue
\mciteSetBstMidEndSepPunct{\mcitedefaultmidpunct}
{\mcitedefaultendpunct}{\mcitedefaultseppunct}\relax
\EndOfBibitem
\bibitem[Pei et~al.(2012)Pei, Laird, Steele, and Kouwenhoven]{Pei:2012jn}
Pei,~F.; Laird,~E.~A.; Steele,~G.~A.; Kouwenhoven,~L.~P. \emph{Nat.
  Nanotechnol.} \textbf{2012}, \emph{7}, 630--634\relax
\mciteBstWouldAddEndPuncttrue
\mciteSetBstMidEndSepPunct{\mcitedefaultmidpunct}
{\mcitedefaultendpunct}{\mcitedefaultseppunct}\relax
\EndOfBibitem
\bibitem[Laird et~al.(2013)Laird, Pei, and Kouwenhoven]{Laird:2013ez}
Laird,~E.~A.; Pei,~F.; Kouwenhoven,~L.~P. \emph{Nat. Nanotechnol.}
  \textbf{2013}, \emph{8}, 565--568\relax
\mciteBstWouldAddEndPuncttrue
\mciteSetBstMidEndSepPunct{\mcitedefaultmidpunct}
{\mcitedefaultendpunct}{\mcitedefaultseppunct}\relax
\EndOfBibitem
\bibitem[Goswami et~al.(2007)Goswami, Slinker, Friesen, McGuire, Truitt, Tahan,
  Klein, Chu, Mooney, van~der Weide, Joynt, Coppersmith, and
  Eriksson]{Goswami:2006bb}
Goswami,~S.; Slinker,~K.~A.; Friesen,~M.; McGuire,~L.~M.; Truitt,~J.~L.;
  Tahan,~C.; Klein,~L.~J.; Chu,~J.~O.; Mooney,~P.~M.; van~der Weide,~D.~W.;
  Joynt,~R.; Coppersmith,~S.~N.; Eriksson,~M.~A. \emph{Nat. Phys.}
  \textbf{2007}, \emph{3}, 41--45\relax
\mciteBstWouldAddEndPuncttrue
\mciteSetBstMidEndSepPunct{\mcitedefaultmidpunct}
{\mcitedefaultendpunct}{\mcitedefaultseppunct}\relax
\EndOfBibitem
\bibitem[Shaji et~al.(2008)Shaji, Simmons, Thalakulam, Klein, Qin, Luo, Savage,
  Lagally, Rimberg, Joynt, Friesen, Blick, Coppersmith, and
  Eriksson]{Shaji:2008hq}
Shaji,~N.; Simmons,~C.~B.; Thalakulam,~M.; Klein,~L.~J.; Qin,~H.; Luo,~H.;
  Savage,~D.~E.; Lagally,~M.~G.; Rimberg,~A.~J.; Joynt,~R.; Friesen,~M.;
  Blick,~R.~H.; Coppersmith,~S.~N.; Eriksson,~M.~A. \emph{Nat. Phys.}
  \textbf{2008}, \emph{4}, 540--544\relax
\mciteBstWouldAddEndPuncttrue
\mciteSetBstMidEndSepPunct{\mcitedefaultmidpunct}
{\mcitedefaultendpunct}{\mcitedefaultseppunct}\relax
\EndOfBibitem
\bibitem[Maune et~al.(2012)Maune, Borselli, Huang, Ladd, Deelman, Holabird,
  Kiselev, Alvarado-Rodriguez, Ross, Schmitz, Sokolich, Watson, Gyure, and
  Hunter]{Maune:2012iu}
Maune,~B.~M.; Borselli,~M.~G.; Huang,~B.; Ladd,~T.~D.; Deelman,~P.~W.;
  Holabird,~K.~S.; Kiselev,~A.~A.; Alvarado-Rodriguez,~I.; Ross,~R.~S.;
  Schmitz,~A.~E.; Sokolich,~M.; Watson,~C.~A.; Gyure,~M.~F.; Hunter,~A.~T.
  \emph{Nature} \textbf{2012}, \emph{481}, 344--347\relax
\mciteBstWouldAddEndPuncttrue
\mciteSetBstMidEndSepPunct{\mcitedefaultmidpunct}
{\mcitedefaultendpunct}{\mcitedefaultseppunct}\relax
\EndOfBibitem
\bibitem[Shi et~al.(2014)Shi, Simmons, Ward, Prance, Wu, Koh, Gamble, Savage,
  Lagally, and Friesen]{Shi:2013uf}
Shi,~Z.; Simmons,~C.~B.; Ward,~D.~R.; Prance,~J.~R.; Wu,~X.; Koh,~T.~S.;
  Gamble,~J.~K.; Savage,~D.~E.; Lagally,~M.~G.; Friesen,~M. \emph{Nat. Commun.}
  \textbf{2014}, \emph{5}, 3020\relax
\mciteBstWouldAddEndPuncttrue
\mciteSetBstMidEndSepPunct{\mcitedefaultmidpunct}
{\mcitedefaultendpunct}{\mcitedefaultseppunct}\relax
\EndOfBibitem
\bibitem[Pla et~al.(2012)Pla, Tan, Dehollain, Lim, Morton, Jamieson, Dzurak,
  and Morello]{Pla:2012jj}
Pla,~J.~J.; Tan,~K.~Y.; Dehollain,~J.~P.; Lim,~W.~H.; Morton,~J. J.~L.;
  Jamieson,~D.~N.; Dzurak,~A.~S.; Morello,~A. \emph{Nature} \textbf{2012},
  \emph{489}, 541--545\relax
\mciteBstWouldAddEndPuncttrue
\mciteSetBstMidEndSepPunct{\mcitedefaultmidpunct}
{\mcitedefaultendpunct}{\mcitedefaultseppunct}\relax
\EndOfBibitem
\bibitem[Pla et~al.(2013)Pla, Tan, Dehollain, Lim, Morton, Zwanenburg,
  Jamieson, Dzurak, and Morello]{Pla:2013io}
Pla,~J.~J.; Tan,~K.~Y.; Dehollain,~J.~P.; Lim,~W.~H.; Morton,~J. J.~L.;
  Zwanenburg,~F.~A.; Jamieson,~D.~N.; Dzurak,~A.~S.; Morello,~A. \emph{Nature}
  \textbf{2013}, \emph{496}, 334--338\relax
\mciteBstWouldAddEndPuncttrue
\mciteSetBstMidEndSepPunct{\mcitedefaultmidpunct}
{\mcitedefaultendpunct}{\mcitedefaultseppunct}\relax
\EndOfBibitem
\bibitem[Dehollain et~al.(2014)Dehollain, Muhonen, Tan, Jamieson, Dzurak, and
  Morello]{Dehollain:2014wc}
Dehollain,~J.~P.; Muhonen,~J.~T.; Tan,~K.~Y.; Jamieson,~D.~N.; Dzurak,~A.~S.;
  Morello,~A. \emph{arXiv.org} \textbf{2014}, \emph{1402.7148v1}\relax
\mciteBstWouldAddEndPuncttrue
\mciteSetBstMidEndSepPunct{\mcitedefaultmidpunct}
{\mcitedefaultendpunct}{\mcitedefaultseppunct}\relax
\EndOfBibitem
\bibitem[Muhonen et~al.(2014)Muhonen, Dehollain, Laucht, Hudson, Sekiguchi,
  Itoh, Jamieson, McCallum, Dzurak, and Morello]{Muhonen:2014tb}
Muhonen,~J.~T.; Dehollain,~J.~P.; Laucht,~A.; Hudson,~F.~E.; Sekiguchi,~T.;
  Itoh,~K.~M.; Jamieson,~D.~N.; McCallum,~J.~C.; Dzurak,~A.~S.; Morello,~A.
  \emph{arXiv.org} \textbf{2014}, \emph{1402.7140v1}\relax
\mciteBstWouldAddEndPuncttrue
\mciteSetBstMidEndSepPunct{\mcitedefaultmidpunct}
{\mcitedefaultendpunct}{\mcitedefaultseppunct}\relax
\EndOfBibitem
\bibitem[Lu et~al.(2005)Lu, Xiang, Timko, Wu, and Lieber]{Lu:2005jx}
Lu,~W.; Xiang,~J.; Timko,~B.~P.; Wu,~Y.; Lieber,~C.~M. \emph{Proc. Natl. Acad.
  Sci. U. S. A.} \textbf{2005}, \emph{102}, 10046--10051\relax
\mciteBstWouldAddEndPuncttrue
\mciteSetBstMidEndSepPunct{\mcitedefaultmidpunct}
{\mcitedefaultendpunct}{\mcitedefaultseppunct}\relax
\EndOfBibitem
\bibitem[Xiang et~al.(2006)Xiang, Lu, Hu, Wu, Yan, and Lieber]{Xiang:2006gi}
Xiang,~J.; Lu,~W.; Hu,~Y.; Wu,~Y.; Yan,~H.; Lieber,~C.~M. \emph{Nature}
  \textbf{2006}, \emph{441}, 489--493\relax
\mciteBstWouldAddEndPuncttrue
\mciteSetBstMidEndSepPunct{\mcitedefaultmidpunct}
{\mcitedefaultendpunct}{\mcitedefaultseppunct}\relax
\EndOfBibitem
\bibitem[Yan et~al.(2011)Yan, Choe, Nam, Hu, Das, Klemic, Ellenbogen, and
  Lieber]{Yan:2011cm}
Yan,~H.; Choe,~H.~S.; Nam,~S.; Hu,~Y.; Das,~S.; Klemic,~J.~F.;
  Ellenbogen,~J.~C.; Lieber,~C.~M. \emph{Nature} \textbf{2011}, \emph{470},
  240--244\relax
\mciteBstWouldAddEndPuncttrue
\mciteSetBstMidEndSepPunct{\mcitedefaultmidpunct}
{\mcitedefaultendpunct}{\mcitedefaultseppunct}\relax
\EndOfBibitem
\bibitem[Yao et~al.(2013)Yao, Yan, and Lieber]{Yao:2013eo}
Yao,~J.; Yan,~H.; Lieber,~C.~M. \emph{Nat. Nanotechnol.} \textbf{2013},
  \emph{8}, 329--335\relax
\mciteBstWouldAddEndPuncttrue
\mciteSetBstMidEndSepPunct{\mcitedefaultmidpunct}
{\mcitedefaultendpunct}{\mcitedefaultseppunct}\relax
\EndOfBibitem
\bibitem[Yao et~al.(2014)Yao, Yan, Das, Klemic, Ellenbogen, and Lieber]{Yao:uh}
Yao,~J.; Yan,~H.; Das,~S.; Klemic,~J.~F.; Ellenbogen,~J.~C.; Lieber,~C.~M.
  \emph{Proc. Natl. Acad. Sci. U. S. A.} \textbf{2014}, \emph{111},
  2431--2435\relax
\mciteBstWouldAddEndPuncttrue
\mciteSetBstMidEndSepPunct{\mcitedefaultmidpunct}
{\mcitedefaultendpunct}{\mcitedefaultseppunct}\relax
\EndOfBibitem
\bibitem[Hao et~al.(2010)Hao, Tu, Cao, Zhou, Li, Guo, Fung, Ji, Guo, and
  Lu]{Hao:2010di}
Hao,~X.-J.; Tu,~T.; Cao,~G.; Zhou,~C.; Li,~H.-O.; Guo,~G.-C.; Fung,~W.~Y.;
  Ji,~Z.; Guo,~G.-P.; Lu,~W. \emph{Nano Lett.} \textbf{2010}, \emph{10},
  2956--2960\relax
\mciteBstWouldAddEndPuncttrue
\mciteSetBstMidEndSepPunct{\mcitedefaultmidpunct}
{\mcitedefaultendpunct}{\mcitedefaultseppunct}\relax
\EndOfBibitem
\bibitem[Kloeffel et~al.(2011)Kloeffel, Trif, and Loss]{Kloeffel:2011bg}
Kloeffel,~C.; Trif,~M.; Loss,~D. \emph{Phys. Rev. B} \textbf{2011}, \emph{84},
  195314\relax
\mciteBstWouldAddEndPuncttrue
\mciteSetBstMidEndSepPunct{\mcitedefaultmidpunct}
{\mcitedefaultendpunct}{\mcitedefaultseppunct}\relax
\EndOfBibitem
\bibitem[Higginbotham et~al.(2014)Higginbotham, Kuemmeth, Larsen, Fitzpatrick,
  Yao, Yan, Lieber, and Marcus]{Higginbotham:2014ui}
Higginbotham,~A.~P.; Kuemmeth,~F.; Larsen,~T.~W.; Fitzpatrick,~M.; Yao,~J.;
  Yan,~H.; Lieber,~C.~M.; Marcus,~C.~M. \emph{arXiv.org} \textbf{2014},
  \emph{1401.2948v1}\relax
\mciteBstWouldAddEndPuncttrue
\mciteSetBstMidEndSepPunct{\mcitedefaultmidpunct}
{\mcitedefaultendpunct}{\mcitedefaultseppunct}\relax
\EndOfBibitem
\bibitem[Nowack et~al.(2007)Nowack, Koppens, Nazarov, and
  Vandersypen]{Nowack:2007du}
Nowack,~K.~C.; Koppens,~F. H.~L.; Nazarov,~Y.~V.; Vandersypen,~L. M.~K.
  \emph{Science} \textbf{2007}, \emph{318}, 1430--1433\relax
\mciteBstWouldAddEndPuncttrue
\mciteSetBstMidEndSepPunct{\mcitedefaultmidpunct}
{\mcitedefaultendpunct}{\mcitedefaultseppunct}\relax
\EndOfBibitem
\bibitem[Nadj-Perge et~al.(2010)Nadj-Perge, Frolov, Bakkers, and
  Kouwenhoven]{NadjPerge:2010kw}
Nadj-Perge,~S.; Frolov,~S.~M.; Bakkers,~E. P. A.~M.; Kouwenhoven,~L.~P.
  \emph{Nature} \textbf{2010}, \emph{468}, 1084--1087\relax
\mciteBstWouldAddEndPuncttrue
\mciteSetBstMidEndSepPunct{\mcitedefaultmidpunct}
{\mcitedefaultendpunct}{\mcitedefaultseppunct}\relax
\EndOfBibitem
\bibitem[Nowack et~al.(2011)Nowack, Shafiei, Laforest, Prawiroatmodjo,
  Schreiber, Reichl, Wegscheider, and Vandersypen]{Nowack:2011cu}
Nowack,~K.~C.; Shafiei,~M.; Laforest,~M.; Prawiroatmodjo,~G. E. D.~K.;
  Schreiber,~L.~R.; Reichl,~C.; Wegscheider,~W.; Vandersypen,~L. M.~K.
  \emph{Science} \textbf{2011}, \emph{333}, 1269--1272\relax
\mciteBstWouldAddEndPuncttrue
\mciteSetBstMidEndSepPunct{\mcitedefaultmidpunct}
{\mcitedefaultendpunct}{\mcitedefaultseppunct}\relax
\EndOfBibitem
\bibitem[van~den Berg et~al.(2013)van~den Berg, Nadj-Perge, Pribiag, Plissard,
  Bakkers, Frolov, and Kouwenhoven]{vandenBerg:2013dk}
van~den Berg,~J. W.~G.; Nadj-Perge,~S.; Pribiag,~V.~S.; Plissard,~S.~R.;
  Bakkers,~E. P. A.~M.; Frolov,~S.~M.; Kouwenhoven,~L.~P. \emph{Phys. Rev.
  Lett.} \textbf{2013}, \emph{110}, 066806\relax
\mciteBstWouldAddEndPuncttrue
\mciteSetBstMidEndSepPunct{\mcitedefaultmidpunct}
{\mcitedefaultendpunct}{\mcitedefaultseppunct}\relax
\EndOfBibitem
\bibitem[Fischer et~al.(2008)Fischer, Coish, Bulaev, and Loss]{Fischer:2008hh}
Fischer,~J.; Coish,~W.~A.; Bulaev,~D.~V.; Loss,~D. \emph{Phys. Rev. B}
  \textbf{2008}, \emph{78}, 155329\relax
\mciteBstWouldAddEndPuncttrue
\mciteSetBstMidEndSepPunct{\mcitedefaultmidpunct}
{\mcitedefaultendpunct}{\mcitedefaultseppunct}\relax
\EndOfBibitem
\bibitem[Pribiag et~al.(2013)Pribiag, Nadj-Perge, Frolov, van~den Berg, van
  Weperen, Plissard, Bakkers, and Kouwenhoven]{Pribiag:2013if}
Pribiag,~V.~S.; Nadj-Perge,~S.; Frolov,~S.~M.; van~den Berg,~J. W.~G.; van
  Weperen,~I.; Plissard,~S.~R.; Bakkers,~E. P. A.~M.; Kouwenhoven,~L.~P.
  \emph{Nat. Nanotechnol.} \textbf{2013}, \emph{8}, 170--174\relax
\mciteBstWouldAddEndPuncttrue
\mciteSetBstMidEndSepPunct{\mcitedefaultmidpunct}
{\mcitedefaultendpunct}{\mcitedefaultseppunct}\relax
\EndOfBibitem
\bibitem[Kloeffel et~al.(2013)Kloeffel, Trif, Stano, and Loss]{Kloeffel:2013iv}
Kloeffel,~C.; Trif,~M.; Stano,~P.; Loss,~D. \emph{Phys. Rev. B} \textbf{2013},
  \emph{88}, 241405\relax
\mciteBstWouldAddEndPuncttrue
\mciteSetBstMidEndSepPunct{\mcitedefaultmidpunct}
{\mcitedefaultendpunct}{\mcitedefaultseppunct}\relax
\EndOfBibitem
\bibitem[Hu et~al.(2007)Hu, Churchill, Reilly, Xiang, Lieber, and
  Marcus]{Hu:2007hu}
Hu,~Y.; Churchill,~H. O.~H.; Reilly,~D.~J.; Xiang,~J.; Lieber,~C.~M.;
  Marcus,~C.~M. \emph{Nat. Nanotechnol.} \textbf{2007}, \emph{2},
  622--625\relax
\mciteBstWouldAddEndPuncttrue
\mciteSetBstMidEndSepPunct{\mcitedefaultmidpunct}
{\mcitedefaultendpunct}{\mcitedefaultseppunct}\relax
\EndOfBibitem
\bibitem[Hu et~al.(2012)Hu, Kuemmeth, Lieber, and Marcus]{Hu:2011ic}
Hu,~Y.; Kuemmeth,~F.; Lieber,~C.~M.; Marcus,~C.~M. \emph{Nat. Nanotechnol.}
  \textbf{2012}, \emph{7}, 47--50\relax
\mciteBstWouldAddEndPuncttrue
\mciteSetBstMidEndSepPunct{\mcitedefaultmidpunct}
{\mcitedefaultendpunct}{\mcitedefaultseppunct}\relax
\EndOfBibitem
\bibitem[Petersson et~al.(2010)Petersson, Smith, Anderson, Atkinson, Jones, and
  Ritchie]{Petersson:2010tq}
Petersson,~K.~D.; Smith,~C.~G.; Anderson,~D.; Atkinson,~P.; Jones,~G.;
  Ritchie,~D.~A. \emph{Nano Lett.} \textbf{2010}, \emph{10}, 2789--2793\relax
\mciteBstWouldAddEndPuncttrue
\mciteSetBstMidEndSepPunct{\mcitedefaultmidpunct}
{\mcitedefaultendpunct}{\mcitedefaultseppunct}\relax
\EndOfBibitem
\bibitem[Jung et~al.(2012)Jung, Schroer, Petersson, and Petta]{Jung:2012dz}
Jung,~M.; Schroer,~M.~D.; Petersson,~K.~D.; Petta,~J.~R. \emph{Appl. Phys.
  Lett.} \textbf{2012}, \emph{100}, 253508\relax
\mciteBstWouldAddEndPuncttrue
\mciteSetBstMidEndSepPunct{\mcitedefaultmidpunct}
{\mcitedefaultendpunct}{\mcitedefaultseppunct}\relax
\EndOfBibitem
\bibitem[Not()]{Note-1}
S. Weinreb LNA SN68. Minicircuits ZP-2MH mixers. Tektronix AWG5014 waveform
  generator used on $V_\mathrm{L}$ and $V_\mathrm{R}$. Coilcraft 0603CS chip
  inductor.\relax
\mciteBstWouldAddEndPunctfalse
\mciteSetBstMidEndSepPunct{\mcitedefaultmidpunct}
{}{\mcitedefaultseppunct}\relax
\EndOfBibitem
\bibitem[van~der Wiel et~al.(2002)van~der Wiel, De~Franceschi, Elzerman,
  Fujisawa, Tarucha, and Kouwenhoven]{vanderWiel:2002gr}
van~der Wiel,~W.; De~Franceschi,~S.; Elzerman,~J.; Fujisawa,~T.; Tarucha,~S.;
  Kouwenhoven,~L. \emph{Rev. Mod. Phys.} \textbf{2002}, \emph{75}, 1--22\relax
\mciteBstWouldAddEndPuncttrue
\mciteSetBstMidEndSepPunct{\mcitedefaultmidpunct}
{\mcitedefaultendpunct}{\mcitedefaultseppunct}\relax
\EndOfBibitem
\bibitem[Higginbotham et~al.(2014)Higginbotham, Kuemmeth, Hanson, Gossard, and
  Marcus]{Higginbotham:2014cg}
Higginbotham,~A.~P.; Kuemmeth,~F.; Hanson,~M.~P.; Gossard,~A.~C.; Marcus,~C.~M.
  \emph{Phys. Rev. Lett.} \textbf{2014}, \emph{112}, 026801\relax
\mciteBstWouldAddEndPuncttrue
\mciteSetBstMidEndSepPunct{\mcitedefaultmidpunct}
{\mcitedefaultendpunct}{\mcitedefaultseppunct}\relax
\EndOfBibitem
\bibitem[Vorojtsov et~al.(2004)Vorojtsov, Mucciolo, and
  Baranger]{Vorojtsov:2004ho}
Vorojtsov,~S.; Mucciolo,~E.~R.; Baranger,~H.~U. \emph{Phys. Rev. B}
  \textbf{2004}, \emph{69}, 115329\relax
\mciteBstWouldAddEndPuncttrue
\mciteSetBstMidEndSepPunct{\mcitedefaultmidpunct}
{\mcitedefaultendpunct}{\mcitedefaultseppunct}\relax
\EndOfBibitem
\bibitem[Johnson et~al.(2005)Johnson, Petta, Marcus, Hanson, and
  Gossard]{Johnson:2005cv}
Johnson,~A.~C.; Petta,~J.~R.; Marcus,~C.~M.; Hanson,~M.~P.; Gossard,~A.~C.
  \emph{Phys. Rev. B} \textbf{2005}, \emph{72}, 165308\relax
\mciteBstWouldAddEndPuncttrue
\mciteSetBstMidEndSepPunct{\mcitedefaultmidpunct}
{\mcitedefaultendpunct}{\mcitedefaultseppunct}\relax
\EndOfBibitem
\bibitem[Danon and Nazarov(2009)Danon, and Nazarov]{Danon:2009id}
Danon,~J.; Nazarov,~Y.~V. \emph{Phys. Rev. B} \textbf{2009}, \emph{80},
  041301\relax
\mciteBstWouldAddEndPuncttrue
\mciteSetBstMidEndSepPunct{\mcitedefaultmidpunct}
{\mcitedefaultendpunct}{\mcitedefaultseppunct}\relax
\EndOfBibitem
\bibitem[Johnson et~al.(2005)Johnson, Petta, Taylor, Yacoby, Lukin, Marcus,
  Hanson, and Gossard]{Johnson:2005ks}
Johnson,~A.~C.; Petta,~J.~R.; Taylor,~J.~M.; Yacoby,~A.; Lukin,~M.~D.;
  Marcus,~C.~M.; Hanson,~M.~P.; Gossard,~A.~C. \emph{Nature} \textbf{2005},
  \emph{435}, 925--928\relax
\mciteBstWouldAddEndPuncttrue
\mciteSetBstMidEndSepPunct{\mcitedefaultmidpunct}
{\mcitedefaultendpunct}{\mcitedefaultseppunct}\relax
\EndOfBibitem
\bibitem[Petersson et~al.(2012)Petersson, McFaul, Schroer, Jung, Taylor, Houck,
  and Petta]{Petersson:2013cv}
Petersson,~K.~D.; McFaul,~L.~W.; Schroer,~M.~D.; Jung,~M.; Taylor,~J.~M.;
  Houck,~A.~A.; Petta,~J.~R. \emph{Nature} \textbf{2012}, \emph{490},
  380--383\relax
\mciteBstWouldAddEndPuncttrue
\mciteSetBstMidEndSepPunct{\mcitedefaultmidpunct}
{\mcitedefaultendpunct}{\mcitedefaultseppunct}\relax
\EndOfBibitem
\bibitem[Maier et~al.(2013)Maier, Kloeffel, and Loss]{Maier:2013gg}
Maier,~F.; Kloeffel,~C.; Loss,~D. \emph{Phys. Rev. B} \textbf{2013}, \emph{87},
  161305\relax
\mciteBstWouldAddEndPuncttrue
\mciteSetBstMidEndSepPunct{\mcitedefaultmidpunct}
{\mcitedefaultendpunct}{\mcitedefaultseppunct}\relax
\EndOfBibitem
\bibitem[Coish and Loss(2005)Coish, and Loss]{Coish:839687}
Coish,~W.~A.; Loss,~D. \emph{Phys. Rev. B} \textbf{2005}, \emph{72},
  125337\relax
\mciteBstWouldAddEndPuncttrue
\mciteSetBstMidEndSepPunct{\mcitedefaultmidpunct}
{\mcitedefaultendpunct}{\mcitedefaultseppunct}\relax
\EndOfBibitem
\bibitem[Cywi{\'n}ski et~al.(2008)Cywi{\'n}ski, Lutchyn, Nave, and
  Das~Sarma]{Cywinski:2008iy}
Cywi{\'n}ski,~{\L}.; Lutchyn,~R.; Nave,~C.; Das~Sarma,~S. \emph{Phys. Rev. B}
  \textbf{2008}, \emph{77}, 174509\relax
\mciteBstWouldAddEndPuncttrue
\mciteSetBstMidEndSepPunct{\mcitedefaultmidpunct}
{\mcitedefaultendpunct}{\mcitedefaultseppunct}\relax
\EndOfBibitem
\bibitem[Beaudoin and Coish(2013)Beaudoin, and Coish]{Beaudoin:2013bm}
Beaudoin,~F.; Coish,~W.~A. \emph{Phys. Rev. B} \textbf{2013}, \emph{88},
  085320\relax
\mciteBstWouldAddEndPuncttrue
\mciteSetBstMidEndSepPunct{\mcitedefaultmidpunct}
{\mcitedefaultendpunct}{\mcitedefaultseppunct}\relax
\EndOfBibitem
\bibitem[Huang and Hu(2013)Huang, and Hu]{Huang:2013wz}
Huang,~P.; Hu,~X. \emph{arXiv.org} \textbf{2013}, \emph{1308.0352v2}\relax
\mciteBstWouldAddEndPuncttrue
\mciteSetBstMidEndSepPunct{\mcitedefaultmidpunct}
{\mcitedefaultendpunct}{\mcitedefaultseppunct}\relax
\EndOfBibitem
\bibitem[Schulten and Wolynes(1978)Schulten, and Wolynes]{Schulten:1978jm}
Schulten,~K.; Wolynes,~P.~G. \emph{J. Chem. Phys.} \textbf{1978}, \emph{68},
  3292--3297\relax
\mciteBstWouldAddEndPuncttrue
\mciteSetBstMidEndSepPunct{\mcitedefaultmidpunct}
{\mcitedefaultendpunct}{\mcitedefaultseppunct}\relax
\EndOfBibitem
\bibitem[Merkulov et~al.(2002)Merkulov, Efros, and Rosen]{Merkulov:2002dz}
Merkulov,~I.~A.; Efros,~A.~L.; Rosen,~M. \emph{Phys. Rev. B} \textbf{2002},
  \emph{65}, 205309\relax
\mciteBstWouldAddEndPuncttrue
\mciteSetBstMidEndSepPunct{\mcitedefaultmidpunct}
{\mcitedefaultendpunct}{\mcitedefaultseppunct}\relax
\EndOfBibitem
\bibitem[Zhang et~al.(2006)Zhang, Dobrovitski, Al-Hassanieh, Dagotto, and
  Harmon]{Zhang:2006bd}
Zhang,~W.; Dobrovitski,~V.~V.; Al-Hassanieh,~K.~A.; Dagotto,~E.; Harmon,~B.~N.
  \emph{Phys. Rev. B} \textbf{2006}, \emph{74}, 205313\relax
\mciteBstWouldAddEndPuncttrue
\mciteSetBstMidEndSepPunct{\mcitedefaultmidpunct}
{\mcitedefaultendpunct}{\mcitedefaultseppunct}\relax
\EndOfBibitem
\bibitem[Not()]{Note-2}
We do not correct for $T_1$ effects in Eqs. (3,4), as $V_\mathrm{max}^{(S,T)}$
  differed significantly from those observed in Fig.~3.\relax
\mciteBstWouldAddEndPunctfalse
\mciteSetBstMidEndSepPunct{\mcitedefaultmidpunct}
{}{\mcitedefaultseppunct}\relax
\EndOfBibitem
\bibitem[Trifunovic et~al.(2012)Trifunovic, Dial, Trif, Wootton, Abebe, Yacoby,
  and Loss]{Trifunovic:2012iq}
Trifunovic,~L.; Dial,~O.; Trif,~M.; Wootton,~J.~R.; Abebe,~R.; Yacoby,~A.;
  Loss,~D. \emph{Phys. Rev. X} \textbf{2012}, \emph{2}, 011006\relax
\mciteBstWouldAddEndPuncttrue
\mciteSetBstMidEndSepPunct{\mcitedefaultmidpunct}
{\mcitedefaultendpunct}{\mcitedefaultseppunct}\relax
\EndOfBibitem
\bibitem[Shulman et~al.(2012)Shulman, Dial, Harvey, Bluhm, Umansky, and
  Yacoby]{Shulman:2012fka}
Shulman,~M.~D.; Dial,~O.~E.; Harvey,~S.~P.; Bluhm,~H.; Umansky,~V.; Yacoby,~A.
  \emph{Science} \textbf{2012}, \emph{336}, 202--205\relax
\mciteBstWouldAddEndPuncttrue
\mciteSetBstMidEndSepPunct{\mcitedefaultmidpunct}
{\mcitedefaultendpunct}{\mcitedefaultseppunct}\relax
\EndOfBibitem
\end{mcitethebibliography}
\end{document}